\documentclass[twocolumn,showpacs,preprintnumbers,amsmath,amssymb]{revtex4}

\usepackage{graphicx}
\usepackage{dcolumn}
\usepackage{bm}
\usepackage[usenames]{color}

\begin{document}


\title{Magnetic anisotropy  in Li-phosphates and
origin of magnetoelectricity in LiNiPO$_{4}$}

\author{Kunihiko Yamauchi}
\author{Silvia Picozzi}%
 \email{silvia.picozzi@aquila.infn.it}
\affiliation{%
Consiglio Nazionale delle Ricerche - Istituto Nazionale di Fisica della Materia (CNR-INFM), CASTI Regional Lab., 67100 L'Aquila, Italy\\
}%

\date{\today}
\newcommand{\lcpo}{LiCoPO$_{4}$ }
\newcommand{\lnpo}{LiNiPO$_{4}$ }
\begin{abstract}
Li-based phosphates are paradigmatic materials for magnetoelectricity. 
By means of first-principles calculations, we elucidate
the microscopic origin of spin anisotropy and of magnetoelectric effects in LiNiPO$_{4}$.  
The comparison with  LiCoPO$_{4}$ reveals that 
Co-${d}^{7}$ and Ni-${d}^{8}$ electronic clouds show distinct orbital shapes,
which in turn result in an opposite trend of the local spin anisotropy with respect to the surrounding O$_{6}$ cages. 
Due to magnetic anisotropy, the Ni-based phosphate shows a peculiar ``angled-cross" spin ground-state, which is responsible for magnetoelectricity. In this respect, we show that,
under a magnetic field $H_{x}$,  an electronic polarization $P_{z}$ arises, 
with an estimated linear magneto-electric coefficient in good agreement with experiments.
\end{abstract}

\pacs{Valid PACS appear here}
\maketitle
Olivine phosphates LiCoPO$_{4}$ and LiNiPO$_{4}$ are attracting large interests, due to their peculiar magnetoelectric (ME) effect [{\em i.e.} the control of ferroelectric (magnetic) properties via a magnetic (electric) field] as well as their application for electrodes in rechargeable Li batteries\cite{batteryexp, batterydft}. Recently, ferrotoroidic domains and antiferromagnetic domains have been independently observed in LiCoPO$_{4}$ by using second harmonic generation.\cite{akennature}
The toroidal moment, generated by a vortex of magnetic moments, is considered as source of a novel {\em ferroic order}, closely related to ME effects\cite{schmit, ederer} as well as to multiferroicity\cite{multiferro1} ({\em i.e.} coexistence of long-range magnetic and dipolar orders).
In fact, LiCoPO$_{4}$ shows a nonzero linear ME coefficient, $\alpha_{xy}$ and $\alpha_{yx}$, at low temperature, consistent with the toroidal moment $T$ nearly parallel to $z$ axis\cite{ederer}, whereas LiNiPO$_{4}$ shows $\alpha_{xz}$ and $\alpha_{zx}$\cite{marcier}.
Although magnetoelectricity has been macroscopically investigated by means of Landau theory\cite{vaknin,chupis}, its microscopic origin has not been clarified yet nor first principles calculations aimed at investigating ME effects in phosphates exist in the literature. 
In LiCoPO$_{4}$, antiferromagnetic (AFM) spins were found to be along the $b$ axis\cite{santoroCo} or uniformly rotated from this axis by 4.6 $^{\circ}$\cite{vaknin}.
In LiNiPO$_{4}$, it was proposed the collinear AFM spins to lie along the $c$ axis \cite{santoroNi}, but recently a
non-collinear structure in a ``spin-cross" style \cite{chupis, jensen} was suggested, in order to explain the butterfly shape of the ME hystheresis curve\cite{rivera.butterfly}.

In this letter, within Density-functional theory (DFT) we investigate the magnetic anisotropy for Li-phosphates LiTMPO$_4$ (TM = Ni, Co) and we are able to correlate the orbital degree of freedom with the calculated anisotropy and spin-configuration. The discussion on magnetic anisotropy is particularly important for  LiNiPO$_{4}$, where we find,  as  magnetic ground-state, a peculiar spin-cross configuration that leads to magnetoelectricity; finally, we clarify the  microscopic origin of ME effects, our calculated ME coefficient being in quantitative agreement with experiments.

{\em Methodology and structural details.}
DFT simulations were performed using the VASP code \cite{vasp} and the PAW pseudopotentials \cite{paw} within the GGA+$U$ formalism\cite{ldau} ($U$=5 eV and $J$=0 eV for Ni $d$-states). Other values of $U$=2 and 8 eV were also tested. 
The cut-off energy for the plane-wave expansion  of the wave-functions was set to 400 eV and a {\bf k}-point shell of (2, 4, 4) was used for the Brillouin zone integration. Lattice parameters were fixed as experimentally observed\cite{lat_licopo4, lat_linipo4}. 
The internal atomic coordinates were fully optimized in a (fully-compensated) AFM configuration, keeping $S_{1}=S_{2}=-S_{3}=-S_{4}$ configuration. 
In the orthorhombic $Pnma$ ($D_{16}^{2h}$) paramagnetic space group, these four Co/Ni sites are related by eight symmetry operations; all the rotations and mirror reflections accompany translations, so that the Co/Ni sites are slightly deviated from high-symmetry positions.  
Optimized coordinates of the four Co/Ni ions are: Co1/Ni1(1/4+$\epsilon$, 1/4, -$\delta$), Co2/Ni2(3/4+$\epsilon$, 1/4, 1/2+$\delta$), Co3/Ni3(3/4-$\epsilon$, 3/4, $\delta$), Co4/Ni4(1/4-$\epsilon$, 3/4, 1/2-$\delta$), where $\epsilon$=0.0268 and $\delta$=0.0207 for Co, $\epsilon$=0.0249 and $\delta$=0.0152 for Ni. 
The spin-orbit coupling (SOC) term was computed 
self-consistently inside each atomic sphere, with a radius of 1 $\rm\AA$. The electronic polarization $P$
was calculated using the Berry phase method\cite{berry}. 

{\em Magnetic anisotropy: LiCoPO$_4$ vs LiNiPO$_4$. }
\begin{figure}
\centerline{\includegraphics[width=0.8\columnwidth]{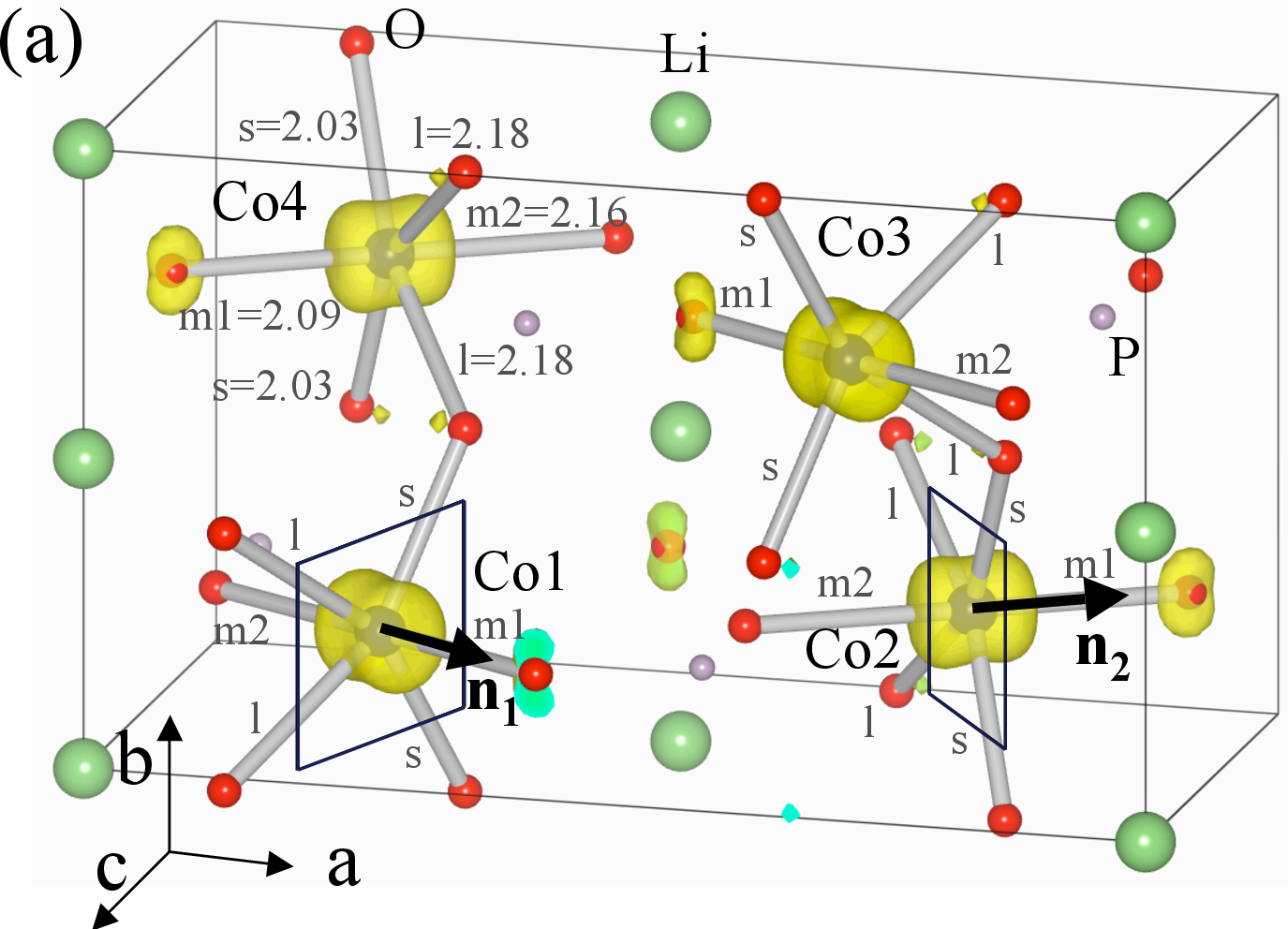}}
\centerline{\includegraphics[width=0.9\columnwidth]{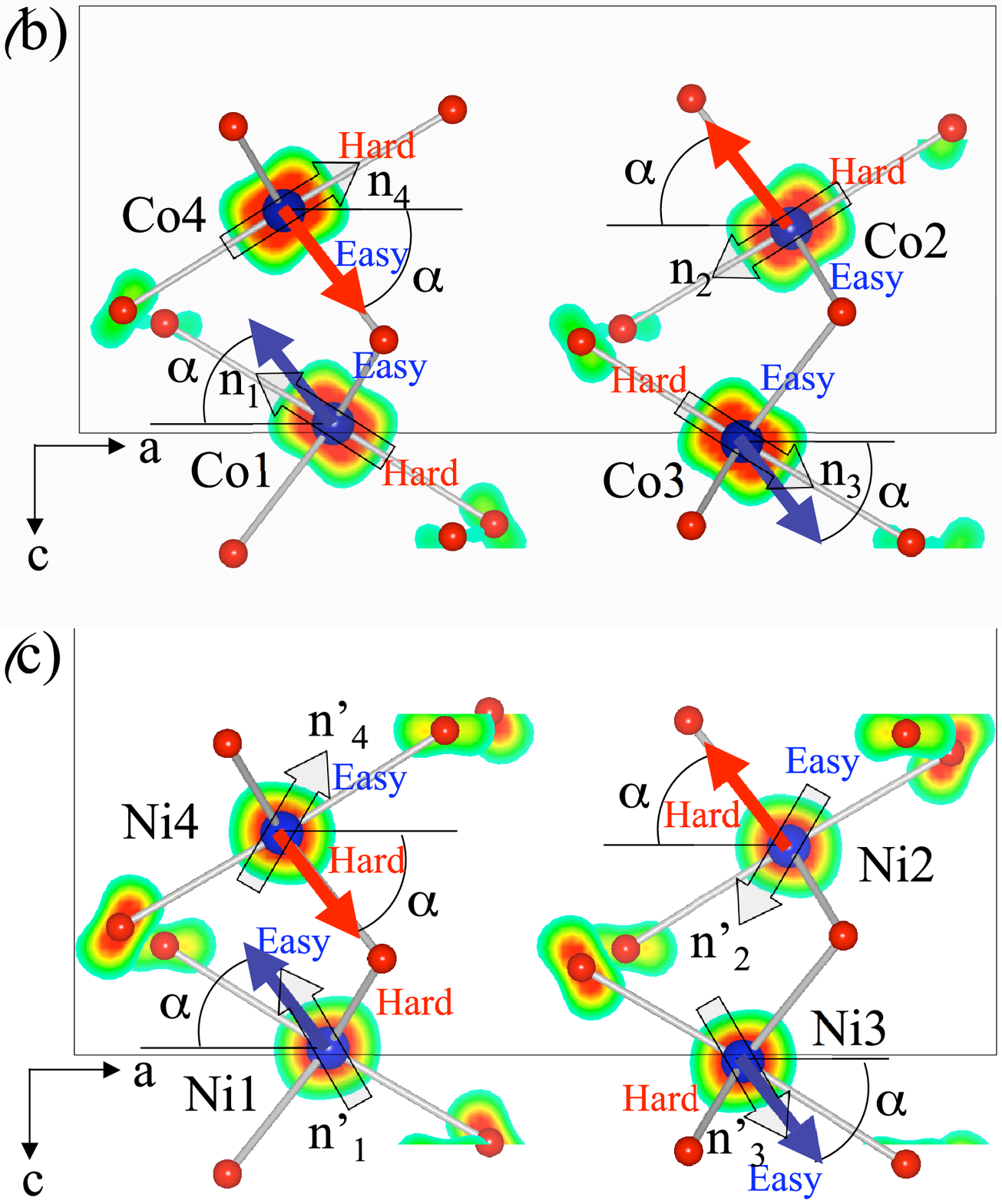}}
\caption{\label{co1to4} 
(a) Isosurface of charge density of minority-spin Co-$t_{2g}$ states (within an energy range  up to 1 eV below $E_{F}$). Co-O bond lenghts (\AA) are reported and denoted as $l, m1, m2, s$. Easy planes for Co local spins and the corresponding normal vectors (hard axis) $\mathbf{n_{i}}$ are also shown at Co1 and Co2 sites. 
(b) Section of charge density of Co $t_{2g}$ electrons and magnetic easy and hard axis in the $ac$ plane  in LiCoPO$_{4}$.  (c) Same as (b) for LiNiPO$_4$. 
Unfilled large arrows show hard ($n_{i}$) and easy ($n'_{i}$) axis in $ac$ plane for Co and Ni spins respectively. 
 } 
\end{figure}
As shown in Fig.\ref{co1to4}, 
each Co ion is surrounded by highly distorted oxygen octahedra, so that the partially filled $t_{2g}$ shell shows a dumbbell-shaped charge distribution, axially elongated along  the $\mathbf{{n_{i}}}$ direction (along m1-m2 O-Co-O bonds). This peculiar shape is expected to induce a strong local magnetic anisotropy via the SOC term. 
On the other hand, in LiNiPO$_{4}$, the
Ni-$d^{8}$ orbital shows a more isotropic sphere--like shape.

Both the global and local magnetic anisotropy were investigated by rotating  four spins simultaneously, keeping a collinear AFM coupling. 
The {\em global} magnetic anisotropic energy (MAE) was calculated by differences in the total energy with different spin-orientations, whereas the {\em local} anisotropic energy was evaluated as proportional to 
the expectation value of the SOC energy: $E_{\rm SOC}=\langle \frac{1}{c^{2}} \frac{1}{r} \frac{dV}{dr} l\cdot s\rangle$\cite{kubler} (integrated in each atomic sphere). 

\begin{figure}
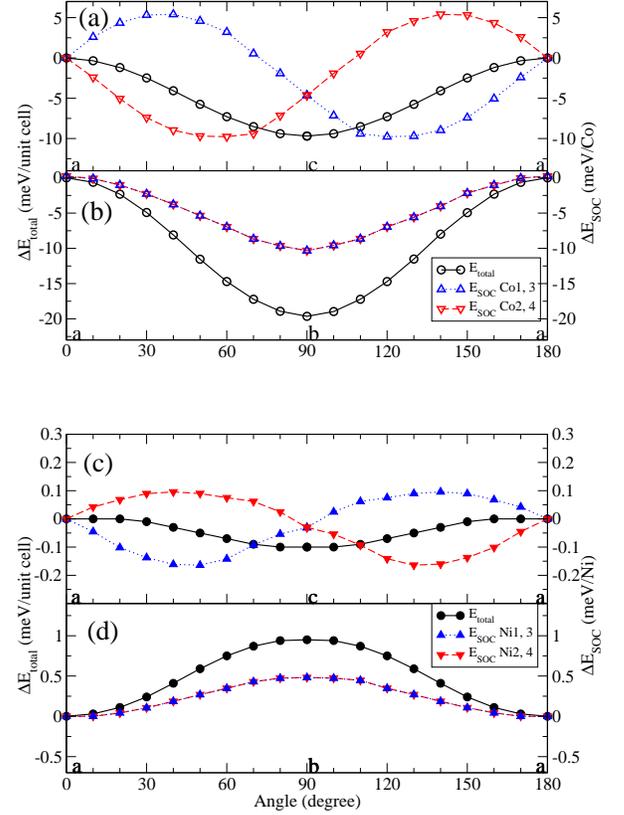



\centerline{\includegraphics[width=0.9\columnwidth]{anisoangCo.eps}}
\vspace{.6cm}
\centerline{\includegraphics[width=0.9\columnwidth]{anisoang.eps}}
\caption{\label{figchgsec} 
Total energy $\Delta E_{total}$ (black solid line) and $\Delta E_{SOC}$ (dashed colored lines) at each Co site vs the direction of collinear Co spins (a) in $ac$ plane and (b) in $ab$ plane. 
Corresponding curves for Ni spins are shown in (c) and (d) for the $ac$ and $ab$ planes, respectively. 
Note the different energy scales of panels (a) and (b) compared to (c) and (d).
The energy with spins along the $a$ axis is taken as reference. 
}
\end{figure}


Fig. \ref{figchgsec} (a) and (b) shows the MAE of Co spins in the $ac$  and $ab$ planes, respectively.
The global easy axis is the $b$ direction, in agreement with the experimental suggestion that spins should be aligned along the $b$ axis (possibly with a slight rotation). The MAE is rather high (more than 10 times larger than the orbitally-ordered Mn spins in TbMnO$_{3}$\cite{xiang.tbmno3}).
As shown by the spin angle dependence of $E_{SOC}$ (cfr Fig. \ref{figchgsec} (a)), the easy/hard axial direction  depends on each Co site.
$E_{tot}$ can be fitted by a conventional quadratic anisotropic term for $S_i$ spins: $D(\mathbf{S_{i}}/|\mathbf{S_{i}}|\cdot\mathbf{n_{i}})^{2}$, where
$\mathbf{n_{i}}$ is the site-dependent hard axis, described as $\mathbf{n_{1}}=\mathbf{n_{3}}=(\cos\alpha,0,\sin\alpha)$, $\mathbf{n_{2}}=\mathbf{n_{4}}=(\cos\alpha,0,-\sin\alpha)$ (see. Fig.\ref{co1to4}). 
According to the fitting,  D=7.39 meV and $\alpha$=35.48$^{\circ}$: this implies the hard axis $\mathbf{n_{i}}$ to be nearly along the $m1$ bond, so that 
Co spins are aligned in an easy plane perpendicular to $\mathbf{n_{i}}$. 
In terms of site-dependent anisotropy, we note that  
 the stable AFM spin ordering $S_{1}=S_{2}=-S_{3}=-S_{4}$ is different from the orbital ordering $L_{1}=L_{3}$, $L_{2}=L_{4}$ which causes a local anisotropy.
In such a configuration, all four spins are allowed to lie in the easy plane {\em only} when collinear AFM spins are pointing along the $b$ axis. 
Therefore the $b$ axis is the easy axis in the collinear AFM configurations for Co spins. 

Our calculations show for Ni spins an opposite trend of $E_{SOC}$, with respect to the Co spin, both in the $ac$ and $ab$ planes (cfr Fig. \ref{figchgsec}); moreover, the MAE is almost two orders of magnitude smaller, consistent with the spherical $d^8$ electronic cloud. 
Here the $b$ axis is the global hard axis.
Assuming the easy axis to lie in the $ac$ plane and by fitting $E_{tot}$ to the anisotropic term, we estimate D=-0.50 meV and $\alpha$=46.7$^{\circ}$ 
Here $\alpha$ is close to $\pi/4$ so that  two opposite contributions from local anisotropy nearly cancel out  (Fig. \ref{figchgsec} (c)) and induce a small in-plane anisotropy. 
The negative value of $D$ implies the existence of an easy axis $\mathbf{n'_{i}}$ of Ni spins in the $ac$ plane
(see. Fig.\ref{co1to4}), $\mathbf{n'_{1}}=\mathbf{n'_{3}}=(-\sin\alpha,0,\cos\alpha)$, $\mathbf{n'_{2}}=\mathbf{n'_{4}}=(\sin\alpha,0,\cos\alpha)$ 
which is close to the $\mathbf{n_{i}}$ direction at Co spins. 
Therefore, the $c$ axis is the global easy axis for Ni spins if a collinear AFM configuration is assumed. However, in order to stabilize the local anisotropic term, the spins are expected to slightly tilt with respect to the $c$ axis. 
This deviation,  denoted as ``angled cross" spin configuration, has been already discussed by Chupis\cite{chupis} in terms of Landau theory  and suggested to play an important role for magnetoelectricity.
\begin{figure}
 \resizebox{80mm}{!}
 {\includegraphics{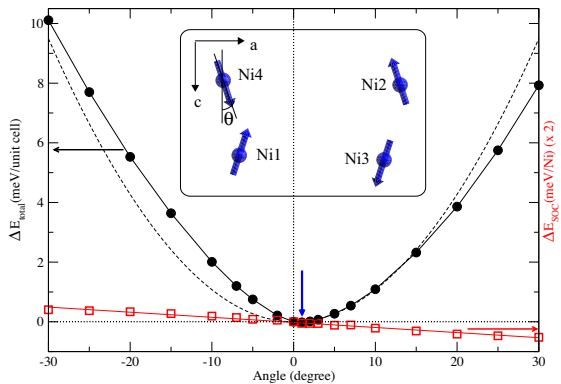}}
 \caption{\label{figcrossen} 
Total energy $\Delta E_{total}$ (black solid line) and $\Delta E_{SOC}$ (dashed red line) at each Ni site vs the direction of non-collinear Ni spins in the ``spin-cross" configuration (shown in the inset).  The minimum is marked by a vertical blue arrow. The dashed line is a function proportional to $1-\cos\theta$, as a guide to the eye, to outline the ``asymmetrical" behaviour of the DFT data.}
\end{figure}
This issue has been carefully investigated here by tilting the spins by an angle $\theta$ from the $c$ axis.
Indeed, as shown in Fig. \ref{figcrossen}, the non-collinear spin structure in the $ac$ plane with $\theta\sim1^{\circ}$ 
gives the lowest energy. 
We  considered a generic Hamiltonian for $\mathbf{S_{i}}$ ($i$ = 1 to 4) spins, including a Heisenberg term, an relativistic anisotropy term and a Zeeman--like term  (see next paragraph, where a finite external H-field will be introduced):
\begin{equation}
\mathcal{H}=\sum_{\langle i,j\rangle}{J_{ij}\frac{\mathbf{S_{i}}\cdot\mathbf{S{_{j}}}}{\left|\mathbf{S_{i}}\right|\left|\mathbf{S_{j}}\right|}}+\sum_{i}{D(\frac{\mathbf{S_{i}}}{\left|\mathbf{S_{i}}\right|}\cdot\mathbf{n_{i}})^{2}} + \sum_{i}{\mathbf{S_{i}}\cdot \mathbf{H}}\\
\label{spinham}
\end{equation}
In this H=0 case, a delicate balance of the first two terms occurs: As the spin is tilted towards the local easy axis, the SOC-term is stabilized, whereas the $J_{ij}$ coupling becomes unstable. As a result, the equilibrium  occurs
in a non-collinear ``angled-cross"  spin-configuration. 

\begin{figure}
\centerline{\includegraphics[width=0.85\columnwidth]{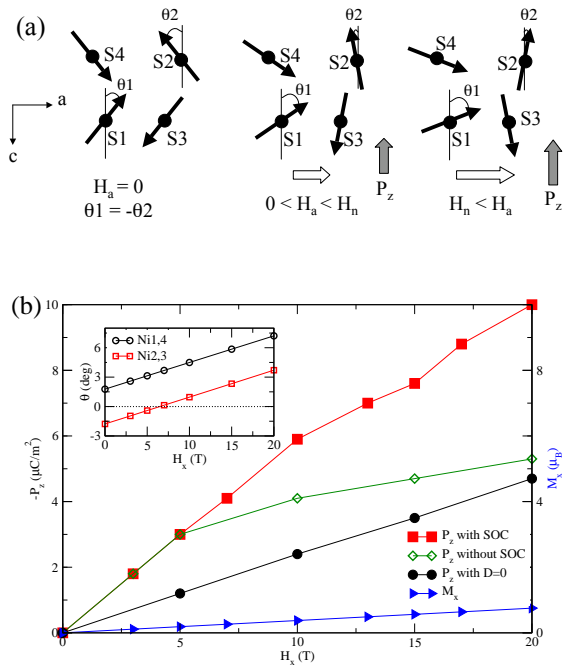}}
\vspace{1.cm}
\centerline{\includegraphics[width=0.85\columnwidth]{pzmx_new.eps}}
\caption{\label{figmxpy} 
(a) Spin configuration in the $ac$  plane under an applied field $H_{x}$. (b) Magnetic-field induced electric polarization along $z$ (with and without SOC, as well as considering D=0 in Eq.(1), see text) in LiNiPO$_4$. The magnetization $M_x$ is also shown (left-trianges, referred to the right $y$-axis) Inset: spin angle $\theta1$ and $\theta2$ (in degrees) vs  $H_{x}$ (T)}.
\end{figure}

{\em Magnetoelectricity in LiNiPO$_4$. }
The existence of a local magnetic anisotropy is particularly important in the context of magnetoelectricity. 
In order to investigate  ME effects in LiNiPO$_{4}$, hereafter we focus on the change in the spin configuration under a magnetic field with respect to the  ``angled cross" ground-state. 
The exchange coupling constants $J_{ij}$ were evaluated by total energy differences fitted 
to Eq. (\ref{spinham}), considering several AFM configurations. Our calculated values are 
 $J_{12}=J_{34}=-0.118  meV, J_{13}=J_{24}=1.46 meV, J_{14}=J_{23}=3.92$meV (the positive sign means AFM coupling).
The dominant $J_{14}$  keeps the spins in the AFM configuration, whereas $J_{12}$ and $J_{13}$ are responsible for removing the degeneracy of some AFM configurations\cite{vaknin}, $i.e.$ the AFM spin configuration $S_{1}$=$S_{2}$=$-S_{3}$=$-S_{4}$ gives the most stable energy. 
The $J$ values are consistent with the fact that four Ni sites are separated into two pairs (Ni1, Ni4) and (Ni2, Ni3) by a Li intercalated layer, so that different pairs are expected to be weakly coupled.
The anisotropy coefficient $D$ is also obtained from DFT, as reported above.
A nonzero applied magnetic field, $H_{x}$, is expected to tilt the Ni spins from the angle-crossed  spin configuration,
with angles from the $c$ axis, denoted as $\theta_{1}$ and $\theta_{2}$.\cite{notaH} 
(Here we assume the relation $\theta_{1}=\theta_{4}$ and $\theta_{2}=\theta_{3}$ by considering the symmetry under the magnetic field.)
By using DFT parameters, Eq.(1) is easily minimized by Newton method with $\theta_{1}$ and $\theta_{2}$ 
angles varying by few degrees in a way proportional to $H_{x}$ (cfr Fig.\ref{figmxpy} b), inset). 

Constraining the 
 spin angles as obtained from the Newton minimization for each H-field, we evaluate the electronic polarization. A small but finite $P_{z}$ is obtained, confirming magnetoelectricity.
As shown in Fig. \ref{figmxpy}, all spins have positive $x$ components  for $H_{x}>$5$T$; however, the sign of the moment doesn't affect the induced $P_{z}$, which is rather linearly proportional to $H_{x}$, consistent with the experimentally observed behaviour at low temperature. 
From the slope of the P-H curve from H=0 to 10 T, the linear ME coefficient $\alpha_{zx}$ is estimated as 0.59 $\mu C/m^{2}T=  0.74 ps/m$, 
comparable with the low-temperature experimental value $\alpha_{zx}^{\rm exp}=1.5 ps/m$.\cite{jensen}

Let us now investigate the
microscopic origin of magnetoelectricity. Electric dipoles are found to be mainly induced by symmetric exchange\cite{prlslv}, often referred to as ``inverse" Goodenough-Kanamori (iGK) interaction \cite{iGK} 
in Ni1-O-Ni4 (or Ni2-O-Ni3) bonds. 
For small $H$, the other conventional mechanism for magnetically--induced polarization  - related to antisymmetric exchange and often labeled as ``inverse Dzyaloshinskii-Moriya (iDM)" -  is negligible. However,  for high magnetic fields, the  iDM becomes sizeable (cfr  Fig.\ref{figmxpy} b), where the comparison between $P_z$ values including - or not - the SOC term is reported).
In this context, we note that the existence of a local anisotropic term is important to induce $P$:
Assuming $D=0$ in Eq.(1), so that $\theta1=\theta2$, we find that the induced $P_{z}$ is less than half with respect to the nonzero $D$ case (cfr Fig.\ref{figmxpy} b).  



Finally, we estimate toroidal moments based on Ref.\cite{ederer}. The collinear AFM spin configuration in LiCoPO$_{4}$ induces (0, 0, -1.52)$\mu_{B}$\AA\ , whereas the noncollinear spin configuration (at H=0T) in LiNiPO$_{4}$ induces (0.19, 0.94, 0) $\mu_{B}$\AA\ per unit cell.
We observe that the spin crossed state causes a small $T_{x}$-component; however the $T_{y}$ component,  coming from the compensated AFM spin configuration, is relevant to the ME effect with finite $P_{z}$ and $H_{x}$. 

In conclusion, we have presented a careful investigation of the spin-anisotropy in Li-based transition-metal phosphates. We
found that LiNiPO$_{4}$ shows a peculiar ground-state spin configuration:  due to a site-dependent magnetic anisotropy, it shows non-collinear spins arranged as ``angled-cross"-like. This is relevant in the context of magnetoelectricity:
due to the ground-state noncollinearity, the spin configuration induced by an applied magnetic field 
leads to a net electric polarization, as shown by a realistic first-principles estimate of the magnetoelectric coefficient in agreement with experiments.
Our results suggest a possible avenue for elucidating the origin of the ME effects in other compounds. 

{\em Acknowledgments.} 
We thank Claude Ederer for helpful discussions.
The research leading to these results has received funding from the European Research Council under the EU Seventh Framework Programme  (FP7/2007-2013) / ERC grant agreement n. 203523.
Computational support from Caspur  Supercomputing Center  (Rome) and Cineca Supercomputing Center (Bologna) is gratefully acknowledged. 



\end{document}